\documentclass[prl, aps, amsfonts, amsmath, amssymb, twoside, twocolumn, superscriptaddress, sort&compress, reprint, 10pt]{revtex4-2}

\usepackage[utf8]{inputenc}
\usepackage[UKenglish]{babel}
\usepackage{nicefrac, graphicx, textcomp, siunitx}
\DeclareSIUnit{\sqrtHz}{\ensuremath{\sqrt{\mathrm{Hz}}}}

\begin{document}

\title{Quantitative imaging of Abrikosov vortices by scanning quantum magnetometry}

\author{Clemens Sch\"afermeier}
\email{clemens.schaefermeier@attocube.com}
\thanks{These authors contributed equally.}
\affiliation{attocube systems GmbH, 85540 Haar, Germany}

\author{Ankit Sharma}
\email{ankit.sharma@attocube.com}
\thanks{These authors contributed equally.}
\affiliation{attocube systems GmbH, 85540 Haar, Germany}

\author{Christopher Kelvin von Grundherr}
\affiliation{attocube systems GmbH, 85540 Haar, Germany}

\author{Dieter Andres}
\affiliation{attocube systems GmbH, 85540 Haar, Germany}

\author{Andrea Morales}
\affiliation{QZabre AG, 8050 Zurich, Switzerland}

\author{Jan Rhensius}
\affiliation{QZabre AG, 8050 Zurich, Switzerland}

\author{Gabriel Puebla-Hellmann}
\affiliation{QZabre AG, 8050 Zurich, Switzerland}

\author{Mirko Bacani}
\affiliation{attocube systems GmbH, 85540 Haar, Germany}

\begin{abstract}
Understanding vortex matter in type-II superconductors is central to controlling dissipation and flux pinning in superconducting materials and devices.
Here, we use cryogenic scanning nitrogen vacancy magnetometry (NVM) to image Abrikosov vortices in the cuprate superconductors BSCCO-2212 and YBCO under controlled field-cooled conditions.
Measurements, which are performed using continuous-wave optically detected magnetic resonance (cw-ODMR) in a closed-cycle cryostat, yield quantitative magnetic-field maps with nanoscale spatial resolution.
In BSCCO-2212 at \SI{71}{\kelvin}, we resolve a well-ordered triangular vortex lattice, whose symmetry and spacing are confirmed through 2D Fourier analysis and are consistent with flux quantization.
YBCO thin films imaged at \SI{3}{\kelvin} exhibit a more disordered vortex arrangement reflecting stronger pinning, while maintaining quantitative agreement between measured vortex density and the applied magnetic field.
These results render our cryogenic scanning NVM a reliable quantitative tool for real-space studies of vortices in high-$T_c$ superconductors, in particular since such a remarkable magnetic resolution has been achieved within relatively short acquisition times of 2 to 4~h.
\end{abstract}

\maketitle

\section{Introduction}

The discovery of high-temperature superconductivity \cite{ref1} in cuprates boosted the research of strongly-correlated electron systems and catalysed progress toward applications in energy technologies and superconducting quantum devices.
The mixed state of type II superconductors is governed by vortex matter---the microscopic manifestation of magnetic flux penetration---whose structure and dynamics are central to dissipation, critical currents, and device stability \cite{ref2}.
When a perpendicular field lies between the lower and upper critical fields, $H_{c1} < H < H_{c2}$, magnetic flux penetrates as quantized Abrikosov vortices.
Each vortex carries one flux quantum, $\Phi_0 = h/2e \simeq \SI{2.07e-15}{\weber}$, with a normal conducting core of size $\xi$ (coherence length) and circulating supercurrents decaying over the London penetration depth $\lambda_L$ \cite{ref2,ref3}.

In clean, weakly pinned materials, repulsive intervortex interactions minimize the Ginzburg--Landau free energy in a triangular (hexagonal) Abrikosov lattice; departures deviations from this ideal structure reflect the interplay of thermal fluctuations, anisotropy, and pinning \cite{ref2}.
This interplay generates a rich phenomenology---vortex solid, vortex glass, and vortex liquid regimes---that directly impacts performance in high-field magnets, superconducting wires, and devices relevant to emerging quantum technologies \cite{ref2,ref3}.

Traditional methods for imaging vortex lattices include Bitter decoration, Lorentz transmission electron microscopy (LTEM), small angle neutron scattering (SANS), muon spin rotation ($\mu$SR), and magneto optical imaging (MOI).
Each has delivered essential insights, but with characteristic limitations: Bitter decoration provides intuitive surface field snapshots yet is effectively destructive/non repeatable on the same region; LTEM requires electron-transparent lamellae and yields projected induction; SANS reports reciprocal-space, volume averaged order; $\mu$SR provides bulk field distributions without direct spatial resolution; and MOI provides large area, video rate maps via an indicator film, at the cost of micrometre scale resolution and a finite stand-off that can perturb or blur fields \cite{ref4,ref5,ref6,ref7}.

Scanning probe approaches such as magnetic force microscopy (MFM) and scanning SQUID have enabled real-space vortex imaging, but MFM can perturb samples via the magnetic tip, and SQUID typically trades spatial resolution for sensitivity and requires cryogenic sensors.
Scanning Hall probe microscopy (SHPM) is quantitative and non contact yet is often limited to \SIrange{.5}{1}{\micro\meter} spatial resolution by sensor size and stand-off; and STM/STS images the quasiparticle DOS in vortex cores with $\sim\xi$ resolution, probing electronic structure rather than stray magnetic field and requiring conductive, atomically prepared surfaces \cite{ref8,ref9,ref10,ref11,ref12,ref13}.

Scanning NVM has emerged as a powerful complementary approach, combining nanoscale spatial resolution with a direct, quantitative conversion between the NV electron spin resonance frequency and the local magnetic field via the Zeeman effect.
In contrast to many other scanning probes, the magnetic field is obtained on an absolute scale without the need for sensor-specific calibration factors.
This makes NV magnetometry NVM particularly well suited for quantitative vortex imaging, where both the number and spacing of vortices can be directly related to flux quantization and the applied magnetic field, as demonstrated in earlier studies \cite{ref14,ref15,ref16}.
However, these early realizations relied on liquid-helium cryostats, custom home-made setups and often required microwave (MW) lines patterned on the sample, thus complicating preparation and incurring constraints.

Here we close that gap with a commercial, closed cycle scanning NV microscope (attoNVM) integrated into an ultra low vibration attoDRY2200 cryostat.
The system combines confocal/wide-field optics with tuning-fork AFM for tip--sample control and employs QZabre diamond tips with an on chip MW line, eliminating on-sample MW patterning while enabling low drift, multi-hour scans and vector-field control over \SIrange{1.8}{300}{\kelvin} at \si{\micro\tesla/\sqrtHz}-level sensitivities \cite{ref17}.

\section{Theoretical Background}

In type-II superconductors, a perpendicular magnetic field between $H_{c1}$ and $H_{c2}$ drives the system into the mixed state, where magnetic flux penetrates the material as quantized Abrikosov vortices, each carrying one flux quantum $\Phi_0$.
The mean vortex density is fixed by flux quantization as
\begin{equation}
n_v=\frac{B}{\Phi_0},
\end{equation}
so that for a scan area $A$ the expected vortex number is
\begin{equation}
N_{\mathrm{exp}}=\frac{BA}{\Phi_0}.
\end{equation}

In the absence of strong disorder, minimization of the Ginzburg--Landau free energy yields a triangular Abrikosov lattice as the equilibrium configuration.
For such a lattice with spacing $a$, the geometric vortex-density is
\begin{equation}
n_v=\frac{2}{\sqrt{3} a^2},
\end{equation}
which, combined with flux quantization, gives the standard relations
\begin{equation}
a=\sqrt{\frac{2\Phi_0}{\sqrt{3} B}},\qquad
B=\frac{2}{\sqrt{3}}\frac{\Phi_0}{a^2}.
\end{equation}

The periodic vortex arrangement is conveniently analysed in reciprocal space via the 2D Fourier transform of the measured field map $B_z(x,y)$.
For an ideal triangular lattice, the reciprocal lattice exhibits six first-order Bragg peaks with magnitude
\begin{equation}
|\mathbf{q}_1|=\frac{4\pi}{\sqrt{3} a}.
\end{equation}
forming a hexagon around the origin in $\lvert \tilde{B}_z(\mathbf{q}) \rvert$.
In experiments, one measures the ring radius
\begin{equation}
f=\frac{|\mathbf{q}_1|}{2\pi}
\end{equation}
of the first order ring in the FFT (in \si{\per\micro\meter}) and obtains
\begin{equation}
a=\frac{2}{\sqrt{3} f},\qquad
B=\frac{\sqrt{3}}{2}\Phi_0 f^2.
\end{equation}
thus, linking the FFT analysis directly to the real-space lattice parameter and average induction.
The sharpness and six-fold symmetry of the Bragg peaks provide a quantitative measure of lattice order.
Narrow rings with six Bragg peaks indicate well-ordered hexagonal lattices, whereas broadened, azimuthally smeared rings reflect pinning, anisotropy, or thermal disorder \cite{ref2,ref18,ref19}.

\section{Experimental Methods}

Experiments were performed using a commercial cryogenic nitrogen vacancy scanning magnetometer, attoNVM, developed together by attocube systems and QZabre, and integrated into an attoDRY2200 closed cycle cryostat with a base temperature of \SI{1.8}{\kelvin} \cite{ref20}.
This cryostat is specifically designed for low-vibration operation, utilizing a proprietary ultra-efficient vibration-damping system to isolate the scanning probe from the mechanical noise of the pulse-tube cold head, which is critical for long-duration nanoscale imaging.

\begin{figure}
  \includegraphics[width=\columnwidth]{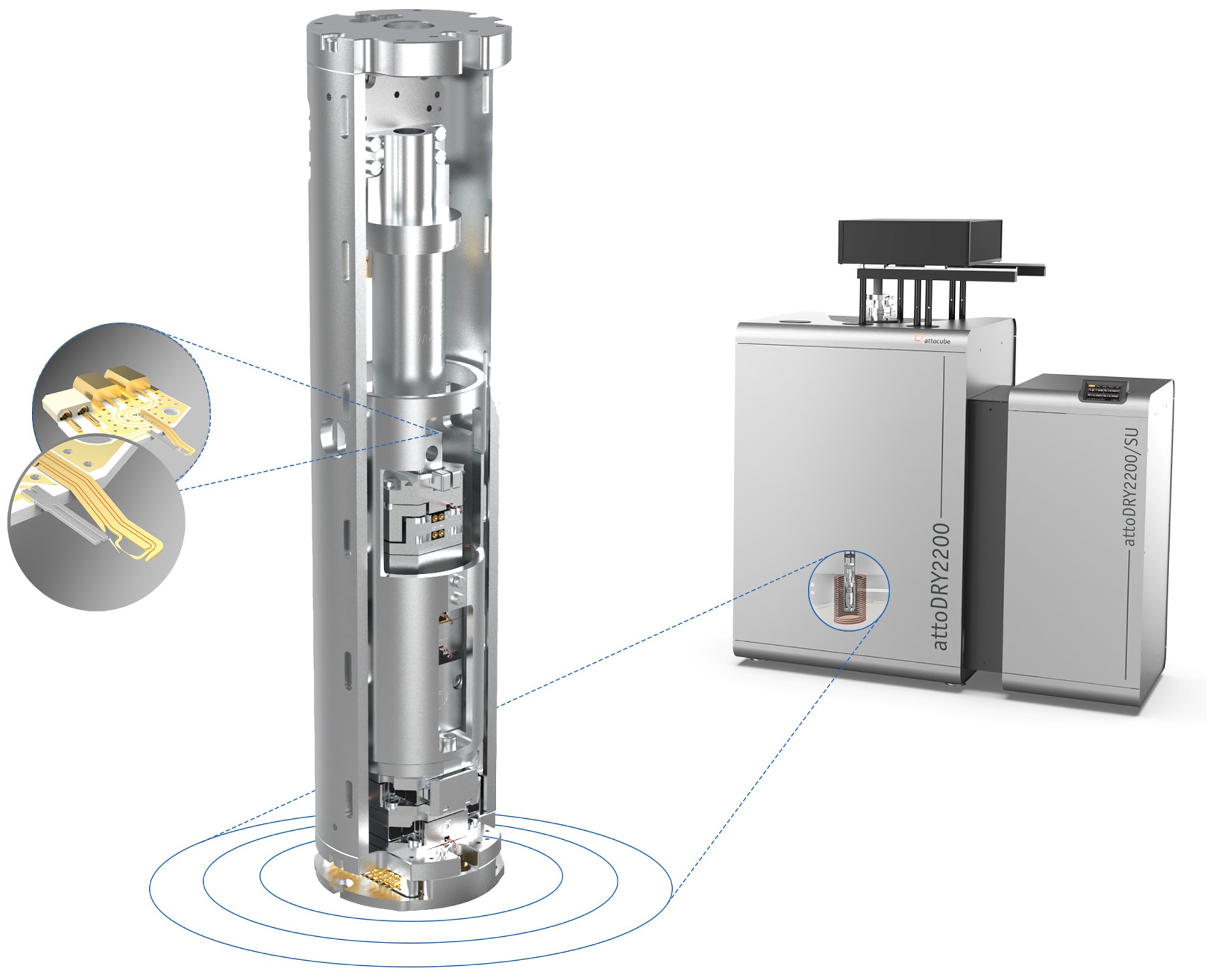}
  \caption{Commercial cryogenic NV magnetometer (attoNVM).
  Schematic overview of the attoNVM microscope integrated into an attoDRY2200 closed cycle cryostat, providing a low vibration, cryogen-free environment for nanoscale magnetic imaging.
  Insets highlight the QZabre diamond probe mounted at the scanner apex, hosting a shallow NV center near the diamond tip apex and an integrated on-chip microwave line on the same carrier chip for efficient ODMR excitation at cryogenic temperatures.}
  \label{fig:setup}
\end{figure}

The microscope combines confocal optical excitation and photoluminescence detection of single NV centers with a wide-field imaging path used for coarse tip--sample alignment.
Fine positioning is achieved with piezoelectric nanopositioners.
Tip--sample distance control is implemented using a quartz tuning fork atomic force microscope operated in shear-force mode.
Depending on the measurement protocol, scans were performed either in fixed height operation or in gentle tip--sample contact; the latter was used for long duration vortex imaging to ensure a stable and reproducible sensor--sample separation.

Magnetic sensing was carried out using a QZabre-fabricated diamond scanning probe hosting a near surface NV center located approximately \SI{10}{\nano\meter} from the tip apex.
The diamond probe and a microwave transmission line are integrated on the same carrier chip and pre-aligned with each other, enabling efficient microwave delivery to the NV center while simplifying probe exchange.
The microwave line is mechanically decoupled from the tuning-fork and cantilever, minimizing parasitic heating and thermal drift at cryogenic temperatures.
No microwave structures on the sample were required.

Magnetic field maps were acquired using cw-ODMR.
At each scan position, optical excitation of the NV center is provided by a \SI{520}{\nano\meter} laser, and the resulting photoluminescence is collected through the same optical path and detected using a single-photon counting module, while the microwave frequency was swept across the lower-frequency NV spin resonance.
The resulting photoluminescence spectra were fitted to extract the resonance frequencies.
The local magnetic field projection along the NV axis was extracted from the shift of the ODMR resonance frequency relative to a reference measurement according to
\[
B_z=(\Delta f-\Delta f_{\mathrm{ref}})/\Upsilon_e
\]
Where $\Delta f$ is the measured resonance frequency shift at each pixel, $\Delta f_{\mathrm{ref}}$ is the reference frequency measured on the sample, and $\Upsilon_e = \SI{28}{\mega\hertz/\milli\tesla}$ is electron gyromagnetic ratio.

Two dimensional magnetic field maps were constructed by repeating this procedure at each pixel of the raster scan.
Typical scans employed a pixel spacing of \SI{66}{\nano\meter} and acquisition times of a few hours, yielding DC magnetic field sensitivities on the order of a few \si{\micro\tesla/\sqrtHz} under cryogenic conditions.

A three-axis vector magnet integrated into the cryostat allows control over both the magnitude and direction of the applied magnetic field.
This capability is used to align the bias field along the NV quantization axis, which maximizes the ODMR Zeeman shift and avoids contrast reduction due to transverse fields.

Two superconducting samples were studied.
A single-crystal Bi$_2$ Sr$_2$CaCu$_2$ O$_{(8+x)}$ (BSCCO-2212) sample was mechanically exfoliated along the ab-plane to expose a clean superconducting surface prior to measurement.
In addition, a \SI{60}{\nano\meter}-thick YBa$_2$ Cu$_3$ O$_{(7-\delta)}$ (YBCO) thin film grown on an Al$_2$ O$_3$ substrate was investigated for comparison.

Vortices were stabilized using a field-cooling procedure.
The samples were cooled from above the superconducting transition temperature to the measurement temperature under a constant magnetic field applied perpendicular to the sample surface.
Under these conditions, the applied net flux is trapped by circulating supercurrents, leading to the formation of Abrikosov vortices pinned by material inhomogeneities.

\section{Results}

\begin{figure}
  \includegraphics[width=\columnwidth]{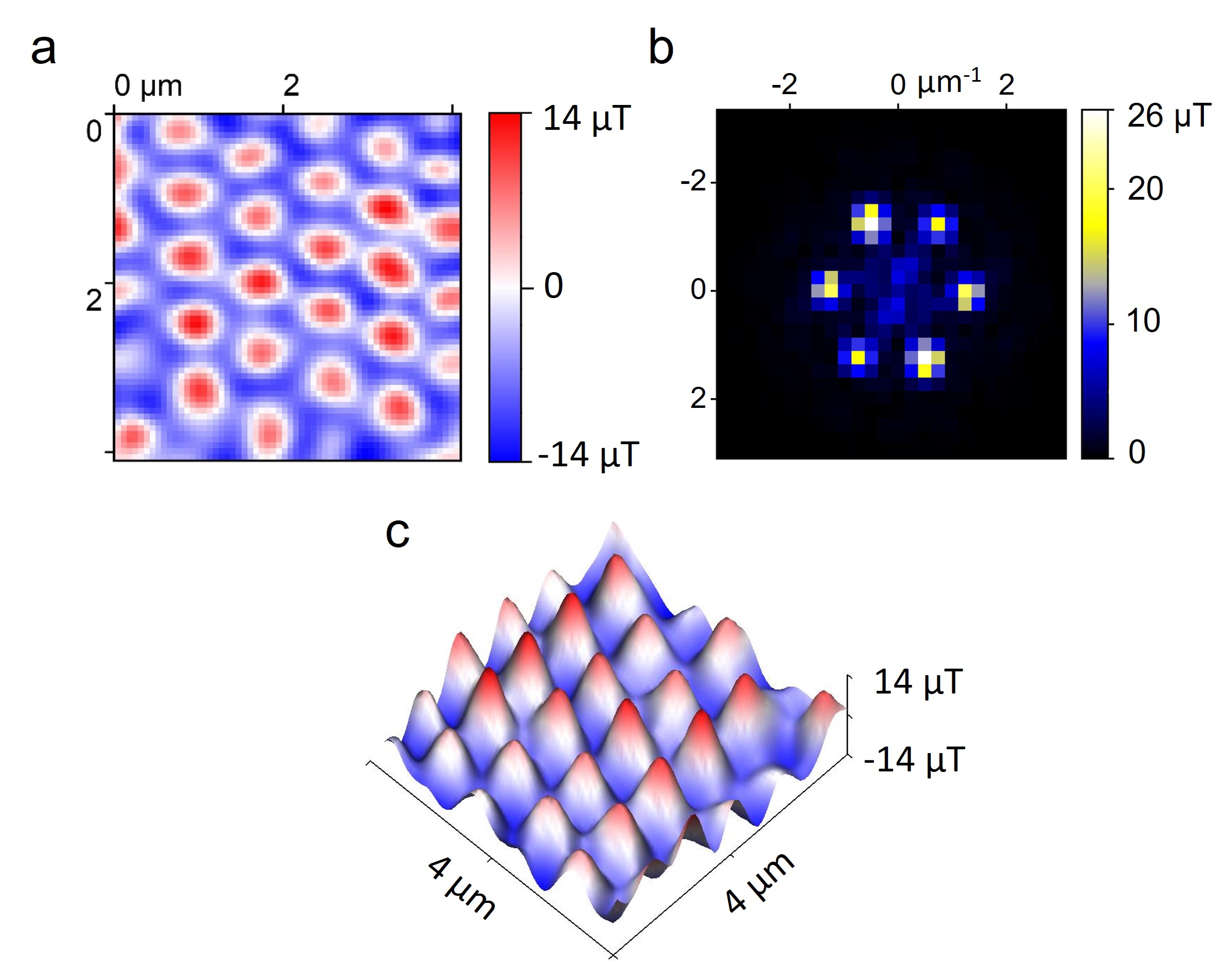}
  \caption{Abrikosov vortices lattice in BSCCO-2212.
  (a) Real-space cw-ODMR magnetic field map acquired in tip--sample contact at $T=\SI{71}{\kelvin}$ after field cooling at $B_z=\SI{3.7}{\milli\tesla}$ and measured under a bias field of \SI{7}{\milli\tesla} projected along the NV axis.
  The pixel spacing is \SI{66}{\nano\meter} and the total acquisition time is 2\,h\,40\,min.
  (b) Two-dimensional fast Fourier transform (FFT) modulus of the magnetic field map in (a).
  The six first-order Bragg peaks reflect the hexagonal symmetry of the triangular Abrikosov vortex lattice.
  (c) Three-dimensional surface rendering of the magnetic field map shown in (a), highlighting the spatial modulation associated with individual vortices.}
  \label{fig:bscco}
\end{figure}

Figure \ref{fig:bscco} shows the magnetic field map of Abrikosov vortices in the BSCCO-2212 single crystal acquired at $T=\SI{71}{\kelvin}$, well below the superconducting transition temperature.
The sample was field-cooled at $B_z = \SI{3.7}{\milli\tesla}$ and imaged under a bias field of $B = \SI{7}{\milli\tesla}$ projected along the NV axis.
The scan area is \SI{4}{\micro\meter^2} with a pixel spacing of \SI{66}{\nano\meter} and a total acquisition time of 2\,h\,40\,min.
The raw magnetic field map (figure \ref{fig:bscco}a) reveals a highly ordered array of localized magnetic field extrema, each corresponding to an individual Abrikosov vortex carrying a single flux quantum $\Phi_0$ \cite{ref2,ref3}.
A total of 26 vortices are observed within the field of view, in good agreement with the expected value of 28.6 vortices calculated from the applied cooling field and scan area using $N = B A / \Phi_0$ \cite{ref2}.

The degree of lattice ordering is more clearly revealed in reciprocal space.
The two-dimensional fast Fourier transform (FFT) of the magnetic field map, shown in figure \ref{fig:bscco}b exhibits a ring of six pronounced maxima corresponding to the first-order Bragg peaks of a triangular vortex lattice.
The presence of these six peaks reflects the hexagonal symmetry and long-range orientational order characteristic of weakly pinned vortex matter in layered cuprate superconductors \cite{ref2,ref19,ref21}.

From the FFT, the radius of the first-order peak ring is measured to be $f = \SI{1.35}{\per\micro\meter}$.
Using the relation for a triangular vortex lattice, $a = 2 / (\sqrt{3} f)$, we obtain a lattice constant $a \approx \SI{0.855}{\micro\meter}$ \cite{ref2,ref18,ref19}.
This spacing corresponds to an effective magnetic field of approximately \SI{3.26}{\milli\tesla}, in close agreement with the applied field-cooling value, demonstrating the quantitative accuracy and internal consistency of the NV-based field measurement.

\begin{figure}
  \includegraphics[width=\columnwidth]{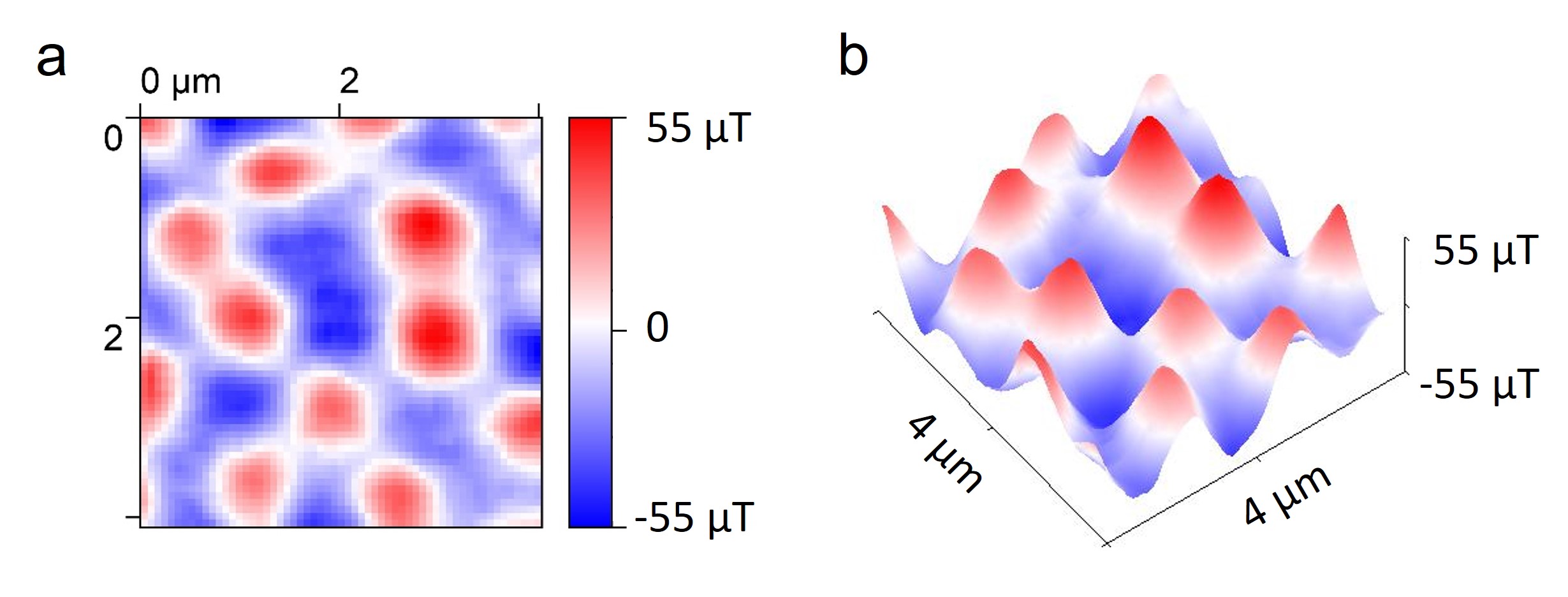}
  \caption{Abrikosov vortices in a YBCO thin film.
  (a) Real-space cw-ODMR magnetic field map acquired in tip--sample contact at $T = \SI{3}{\kelvin}$ after field cooling at $B_z = \SI{1.18}{\milli\tesla}$ and measured under an NV-aligned bias field of \SI{1.84}{\milli\tesla}.
  The pixel spacing is \SI{66}{\nano\meter} and the total acquisition time is 4\,h.
  (b) Three-dimensional surface rendering of the magnetic field map shown in (a), illustrating the spatial variation associated with individual vortices.}
  \label{fig:ybco}
\end{figure}

Figure \ref{fig:ybco} shows the magnetic field map of Abrikosov vortices in the \SI{60}{\nano\meter}-thick YBCO thin film acquired at $T = \SI{3}{\kelvin}$.
The sample was field-cooled at $B_z = \SI{1.18}{\milli\tesla}$ and imaged under a measurement bias field of $B = \SI{1.84}{\milli\tesla}$ projected along the NV axis.
The scan parameters were identical to those used for the BSCCO measurements, with a pixel spacing of \SI{66}{\nano\meter} and a total acquisition time of 4\,h.
Nine vortices are clearly resolved within the scanned area, in excellent agreement with the theoretical expectation of $N_{\mathrm{exp}}=9.14$, confirming the quantitative reliability of vortex counting based on flux quantization \cite{ref2,ref22}.

In contrast to the BSCCO data, the vortices in the YBCO thin film do not form a regular lattice.
While individual vortex cores are clearly resolved, their spatial distribution is irregular.
This difference is reflected in the FFT, which lacks sharp Bragg peaks and instead exhibits diffuse intensity near the expected reciprocal-space positions, indicative of short-range correlations rather than long-range crystalline order \cite{ref19,ref23}.

Such behaviour is consistent with strong pinning and reduced vortex mobility in YBCO thin films at low temperatures, where defects, strain, and film microstructure dominate vortex arrangement \cite{ref2,ref23,ref24}.
Despite the reduced lattice order, the measured vortex density remains consistent with the applied magnetic field, demonstrating that scanning NV magnetometry provides reliable quantitative information even in strongly disordered vortex configurations.

\section{Conclusion}

We have demonstrated quantitative imaging of Abrikosov vortices in two high-$T_c$ superconductors---BSCCO-2212 and a YBCO thin film---using the attoNVM, a cryogenic scanning nitrogen vacancy magnetometer.
All data presented here was acquired within 4 days including cool-down and warm-up of the cryostat, that is, 3 cooling cycles.
This demonstrates the reliable operation of the attoNVM.

Real-space magnetic-field maps acquired via continuous-wave ODMR resolve individual vortices and their spatial ordering under well-defined field-cooled conditions.
For the BSCCO single crystal, a triangular vortex lattice is observed, with lattice spacing and magnetic induction consistent with flux quantization.
In the YBCO thin film, a more disordered vortex arrangement is resolved at low temperature, while the measured vortex density remains quantitatively consistent with the applied magnetic field.

The agreement between vortex densities obtained from direct counting and those extracted from Fourier-space analysis highlights the internal consistency of the measurement approach and the quantitative nature of the attoNVM.
These results render the attoNVM a robust and reproducible tool for studying vortex matter without the need for liquid helium or sample-integrated microwave structures.
The approach is readily extendable to investigations of vortex dynamics, pinning landscapes, and engineered superconducting heterostructures across a broad temperature and magnetic-field range.

\begin{acknowledgments}
The authors would like to thank Andreas Erb for providing us the BSCCO sample.
\end{acknowledgments}

%
\end{document}